\documentclass[twocolumn,showpacs,preprintnumbers,amsmath,amssymb]{revtex4}
\usepackage{amsfonts}

\newcommand{\be}{\begin{equation}}
\newcommand{\ee}{\end{equation}}
\newcommand{\ba}{\begin{eqnarray}}
\newcommand{\ea}{\end{eqnarray}}
\usepackage{graphicx}
\usepackage{dcolumn}
\usepackage{bm}

\begin{document}

\preprint{Physical Review Letters}

\title{A coordination-based approach to elasticity  of floppy and stiff random networks}
\author{ M. Wyart, H. Liang, A. Kabla and L. Mahadevan}
\affiliation{School of Engineering and Applied Sciences, Harvard University, 29 Oxford Street, Cambridge, MA 02138}
\date{\today}

\begin{abstract}
We study  the role of connectivity on the  linear and nonlinear elastic behavior of amorphous systems using a two-dimensional random network of harmonic springs as a model system. A natural characterization of these systems arises in terms of the network  coordination relative to that of an isostatic network $\delta z$; a floppy network has $\delta z<0$, while a stiff network has $\delta z>0$.  Under the influence of an externally applied load we observe that the response of both floppy and rigid network  are controlled by the same critical point, corresponding to the onset of rigidity. We use numerical simulations to compute the exponents which characterize  the shear modulus,  the amplitude of non-affine displacements, and the network stiffening as a function of $\delta z$, derive these theoretically and make predictions for the mechanical response of glasses and fibrous networks.

\end{abstract}


\maketitle

The mechanics of crystalline solids is a fairly well understood subject owing to the simplicity of the underlying lattice which is periodic. In contrast, an understanding of the mechanics of amorphous solids is complicated by the presence of quenched disorder, often on multiple scales. 
Two structural properties affecting the elasticity of disordered solids are their coordination, and the presence of different types of interactions between the constituents of vastly dissimilar strengths. In the case of weakly-coordinated covalent glass such as amorphous selenium, the backbone is floppy, {\it i.e.} it is continuously deformable with almost no energy cost, but weak interactions such as van der Waals are responsible for the non-vanishing elastic moduli. On the other hand, highly-coordinated covalent glasses such as silica, or amorphous particle assemblies where the main interaction is radial, such as emulsions, metallic glasses or granular matter,  the backbone is stiff.  In foams and fibrous networks which are made of low-dimensional structures such as filaments and membranes,  there is a wide separation of energetic scales between  stretching and bending modes. This leads to a range of curious mechanical responses in these materials including strongly non-affine deformations \cite{liu,roux2, emulsion, tanguy, leon,wouter} and  elastic moduli that can be quite sensitive to the applied stress \cite{liu, tanaka}. Despite several theoretical advances \cite{thorpe1, thorpe2, frey, maloney, lubensky}, a unified descriptions of these behaviors remains to be given. Here we study the mechanical response of simple floppy and rigid systems as the coordination is continuously varied and propose such a unifying approach.

\begin{figure}
\includegraphics[width=0.48\textwidth]{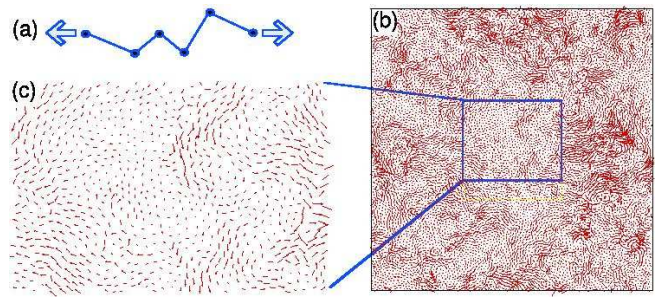}
\caption{\label{fig1} (a) Stretching a zig-zag chain of springs that are freely hinged costs no energy until they are aligned; (b) A random network of such springs, when subjected to shear, exhibits a nonaffine displacement field (see text for definition). Here we show the result for 10000 random springs with coordination $z=z_c=4.0$, subject to a shear strain $\gamma=0.005$; (c) Zooming in shows the presence of large correlated rotational deformations. }
\end{figure}

We start by recalling Maxwell's criterion for rigidity in a central force network \cite{max} by considering a set of $N$ points in $d$ dimensions, subject to $N_c$ constraints in the form of bonds that connect these points. This network has $Nd-N_c$ effective degrees of freedom (ignoring the $d(d+1)/2$ rigid motions of the entire system), and an average coordination number $z=2N_c/N$. The system is said to be isostatic when the system is just rigid, i.e. the number of constraints and the number of degrees of freedom are just balanced, so that $Nd=N_c$, and $z = 2d$. 
When $z<z_c$,  the network exhibits collective degrees of freedom with no restoring force; these solutions are called soft modes. For such a network made of harmonic springs of stiffness $k$, the energy can be written as 
\be \label{1} \delta
E=\sum_{\langle ij\rangle} \frac{k}{2} [(\delta {\vec
R_i}-\delta {\vec R_j})\cdot {\vec n_{ij}}]^2 +o(\delta R^2)
\ee 
where  ${\vec n_{ij}}$ is the unit vector going from $i$ to $j$, and $\delta {\vec R_i}$ is the displacement of particles $i$. Soft modes satisfy $\delta E=0$, or equivalently $(\delta {\vec R_i}-\delta {\vec R_j})\cdot
{\vec n_{ij}}=0 \ \forall ij$. In Fig(\ref{fig1}.a), we show the example of a one dimensional zig-zag structure that straightens without an energy cost until the external load does not couple to the soft modes, {\it i.e.} when the latter correspond to node displacements that are orthogonal to the load direction; the system then becomes stiff. 

To understand how this simple idea carries over to a floppy network, we created disordered two-dimensional networks with up to 10000 particles  using biperiodic jammed configurations of bi-disperse particles \cite{J}.   Linear springs were then used to connect neighboring particles; starting with $z \approx 5.5 > z_c$, a family of networks with lower coordination is then generated by progressively removing links from the most connected pairs of nodes leading to isotropic networks with low heterogeneity in both density and coordination, with a range of spring rest lengths with the ratios of extremes varying by a factor of less than 2. The size of the smallest springs is $l$, and their stiffness $k$. These networks are different from those studied in rigidity percolation \cite{thorpe1} or self-organized networks \cite{thorpe2}, where the fluctuations in coordination are dominant and can rigidify the system even if $z<z_c$. In addition, to model the presence of weak interactions, we introduce weak springs (of dimensionless stiffness $k_w \equiv{\tilde k_w}/k=  10^{-5}$ excepted when stated otherwise. ${\tilde{}}$ designates dimensional  quantities) with a number density $\rho_w$   which stabilize the system.
We impose a pure shear deformation on the  network incrementally, and minimize the system energy via a damped molecular dynamics method. The spatially averaged dimensionless Cauchy stress is then calculated as $\sigma\equiv {\tilde \sigma}/k=\frac{1}{2V}\sum_{ij}{\vec f}_{ij}\bigotimes ({\vec R_j}-{\vec R_i})$, where the sum is on all springs  $ij$, $V={\tilde V}/l^2$ is the dimensionless area of the simulation cell, ${\vec f}_{ij}\equiv {\vec {\tilde f}}_{ij}/kl$ is the dimensionless force vector in the link $ij$, and ${\vec R_i}\equiv {\vec{\tilde R_i}}/l$ is the dimensionless position of particle $i$.  Fig(\ref{fig1}.b,c)  shows the  response to shear for $z=4$ and $\gamma=0.05$. 

\begin{figure}
\includegraphics[width=0.48\textwidth]{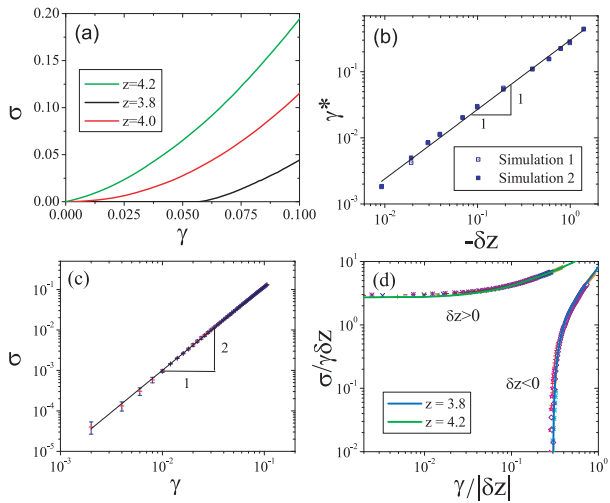}
\caption{\label{fig2} (a) Dimensionless stress-strain curves for $z=3.8, 4.0$, and $4.2$. (b) Critical strain $\gamma ^*$ \emph{vs}. $\delta z\equiv z-4$.  Each point averages over 2 configurations.  (c) Log-log plot of dimensionless stress-strain curve for $z=4$. (d) Rescaled stress-strain curves for $z \in [3.0, 4.99]$  shows that  both floppy and stiff networks can be described in terms of the relative coordination $\delta z$.}
\end{figure}

In  Fig.(\ref{fig2}.a) we quantify the elastic properties as a function of parameters, and show the existence of three qualitatively different stress-strain relations. For floppy networks with $z-z_c=\delta z<0$,  there is a critical strain $\gamma^*$ separating a zero stress plateau and a strain-stiffening regime; this critical strain $\gamma^*$  is a function of the deficit in coordination number $\delta z$ and follows the scaling law  $\gamma^*\sim |\delta z|^\beta$, with $\beta=1$, over nearly two decades up to $z=3$ and a strain as large as $40\%$. For  an isostatic system with $\delta z=0$, the system resists shear deformations nonlinearly as soon as $\gamma \ne 0$; indeed the stress-strain relation is parabolic $\sigma\sim \gamma^2$, as shown in Fig.(\ref{fig2}.c). Finally, for a rigid system with $\delta z>0$, a linear stress-strain regime can be identified; however both the value of the dimensionless shear modulus $G\equiv {\tilde G}/k$ and the extent of the linear regime vanishes as $\delta z\rightarrow 0$, as we discuss quantitatively below.

Since the scaling relations in Fig.~(\ref{fig2}.b,c) are reminiscent of a critical point, we look for scaling functions  on which all the stress-strain
curves must collapse after a suitable re-scaling of the axis. Postulating $G \sim \delta z^\theta$ for $z > z_c$, we  write: 
 \be \label{2}
\sigma= G(\delta z) \gamma f_{\pm}(\gamma/\gamma^*)\equiv  \delta z^\theta \gamma f_{\pm}(\frac{\gamma}{\delta
z^\beta}) 
\ee
 where the functions $f_+$ ($f_-$) characterize the stress-strain relation of rigid (floppy) networks when $\delta z>0$ ($\delta z<0$). When the argument of $f$ vanishes, the existence of both  a linear and floppy regime implies that $f_+(x)\rightarrow c$, where $c$ is a constant of order one, and $f_-(x)\rightarrow0$. For strains much larger than the characteristic strain $\gamma^*$ , the shear modulus must depend on the strain but not depend significantly on the value of $\delta z$. This implies the existence of a power-law form $f(x)\sim x^\chi$, with $\chi=\theta/\beta$. Previous numerical and empirical studies \cite{J,wouter, roux2}  show that $\theta=1$ in two and three dimensions, and were justified theoretically  \cite{thesis}. 
Our numerical results  in Fig.(\ref{fig2}.b) imply that $\beta=1$ and Fig.(\ref{fig2}.c) implies $\chi=1$, in agreement with the relation derived. To directly check the validity of Eq.(\ref{2}) we rescale the axes of Fig.(\ref{fig2}.a), and show the results in Fig.(\ref{fig2}.d)   vindicating our choice of scaling functions.

\begin{figure}
\includegraphics[width=0.5\textwidth]{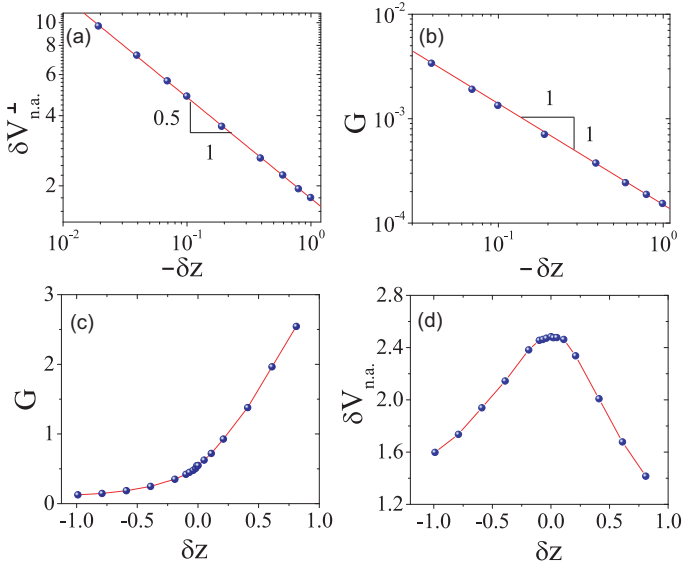}
\caption{\label{fig3} 
(a) The relative non-affine displacement per unit strain as defined in the text $\delta V^\bot_{n.a.}$ for $\gamma = \gamma ^*/10$ and (b) the dimensionless shear modulus $G$   {\it vs}. $-\delta z$, in the floppy regime, for $k_w=10^{-5}$.  The variation in (c) the dimensionless shear modulus $G$ and  (d) the velocity $\delta V^\bot_{n.a.}$ {\it vs.} 
$\delta z \in [-1,1] $, with $k_w=1/300, \rho_w=4.1$. We observe in particular the large particle velocity in the vicinity of the isostatic point $\delta z=0$.}
\end{figure}

Moving beyond the scaling properties of the stress-strain curves, we now consider  the structure of the displacement fields shown in Fig.(\ref{fig1}.b). The non-affine
displacement field $\{ \delta {\vec R^i_{n.a.}(\gamma)} \}_{i=1...N}$ is defined as $ \delta {\vec R^i_{n.a.}}= {\vec R}^i  - {\vec R_{a.}}^i$, where ${\vec R}^i$ is the
final equilibrated configuration, and ${\vec R_{a.}}^i$ is the affine configuration obtained after the homogeneous strain $\gamma$ is imposed. 
We focus on  the amplitude of the {\it relative} displacement among nearest neighbors, defined as $\delta R^\bot_{n.a.}\equiv \langle |\delta {\vec R^i_{n.a.}}- \delta {\vec R^j_{n.a.}}|\rangle$ where the average considers all  pairs of particles in contact. This is a measure of the rotation of individual springs in the non-affine field, and the variation in the relative displacement  per unit strain defined as $\delta V^\bot_{n.a.}\equiv  \partial \delta R^\bot_{n.a.}(\gamma)/ \partial \gamma$ characterizes this rotation as a function of strain increment.  In Fig(\ref{fig3}.a), we show that $\delta V^\bot_{n.a.}\sim |\delta z|^\phi$ with $\phi=-1/2$, i.e. as the rigid sub-skeleton or backbone becomes strongly constrained, larger and larger displacements are required to accommodate a given shear. The rigid backbone acts
as a lever, whose amplification factor diverges as one approaches the onset of rigidity corresponding to $\delta z = 0$.   This scaling has already been observed in
assemblies of elastic particles above the rigidity onset \cite{wouter}.

A small affine deformation of amplitude $\gamma$ causes dimensionless  forces $|\delta F\rangle \equiv \{\vec F^i\}$   on each node. Assuming for convenience of notation that the stiffnesses and lengths of all spring are equal leads to $\vec F^i =\sum_{j} ({\vec n_{ij}}\cdot \Gamma \cdot {\vec n_{ij}}) {\vec n_{ij}}$, where $\Gamma$ is the corresponding strain tensor, and the sum is on all the neighbor $j$ of $i$. For a rigid system with $\delta z>0$, the non-affine displacements $|\delta R\rangle \equiv \{ \delta \vec R^i_{n.a.}\}$, which  corresponds to the displacement of the particle along the direction of unbalanced forces, is:
\be
\label{11}
|\delta R\rangle = M^{-1} |\delta F\rangle = \sum_{\omega} \frac{1}{\omega^2} \langle \delta R(\omega)|\delta F\rangle |\delta R(\omega)\rangle
\ee
where $M$ is the dynamical matrix and $ |\delta R(\omega)\rangle$ is the normalized normal mode of stiffness $\omega^2$. Then the relative displacement between neighboring particles is:
\be
\label{12}
\delta {\vec R_{ij}}\equiv  \delta {\vec R_i}- \delta {\vec R_j}= \sum_{\omega} \frac{1}{\omega^2} \langle \delta R(\omega)|\delta F\rangle \delta {\vec R_{ij}}(\omega)
\ee
Following the justification of \cite{maloney}, we shall assume that the contribution of the different modes are independent:
\be
\label{13}
\langle ||\delta {\vec R_{ij}}||^2\rangle  =  \sum_{\omega} \frac{1}{\omega^4} \langle \delta R(\omega)|\delta F\rangle^2 \langle ||\delta {\vec R_{ij}}(\omega)||^2\rangle
\ee
For weakly coordinated systems, we shall use the results of \cite{matthieu1} which show that above some frequency $\omega^*\sim \delta z$, (i) the density of states $D(\omega)$  does not depend significantly on $\omega$ and (ii) above $\omega^*$, the  modes are very heterogeneous, so that the correlations of the displacements of the particles are weak, see also \cite{sil} .  For such modes $\langle ||\delta {\vec R_{ij}}(\omega)||^2\rangle\sim \langle ||\delta{\vec  R_i}(\omega)||^2\rangle =1/N$, the latter equality stemming from the normalization of the modes. To estimate $\langle \delta R(\omega)|\delta F\rangle^2 = [ \sum_{ij} (\delta {\vec R_i}-\delta {\vec R_j})\cdot {\vec n_{ij}}) ({\vec n_{ij}}\cdot \Gamma \cdot {\vec n_{ij}}) ]^2 $, we use the weak spatial correlation of the modes and treat the different terms as independent. Using  $\langle ({\vec n_{ij}}\cdot \Gamma \cdot {\vec n_{ij}}) ]^2\rangle \sim \gamma^2$ then leads to
\be
\langle \delta R(\omega)|\delta F\rangle^2 \sim \gamma^2 \sum_{ij} [(\delta {\vec R_i(\omega)}-\delta {\vec R_j(\omega)})\cdot {\vec n_{ij}})]^2\sim \gamma^2 \omega^2
\ee
where we used the definition  $\omega^2= 1/2  \sum_{ij} [(\delta {\vec R_i(\omega)}-\delta {\vec R_j(\omega)})\cdot {\vec n_{ij}})]^2$.  Finally, in the large $N$ limit we have $\sum_{\omega} 1/N \rightarrow \int d\omega D(\omega)$, so that Eq.(\ref{13}) yields:
\be
\label{14}
\langle ||\delta {\vec R_{ij}}||^2\rangle/\gamma^2\sim \int d\omega \frac{D(\omega)}{\omega^2} > \int_{\omega>\omega^*} d\omega \frac{1}{\omega^2}\sim \frac{1}{\omega^*}\sim \frac{1}{\delta z}
\ee
 leading to the relation $ \delta R^\bot_{n.a.}\sim ||\delta {\vec R_{ij}}|| \sim \gamma/{\sqrt \delta z}$.

From this we may deduce the strain at which stiffening occurs. The  small applied affine deformation $\gamma$  causes forces ${\vec F^i} \sim \gamma$, which lead to small non-affine displacements  $\delta {\vec R_{n.a.}^i}\approx \gamma {\vec u^i}$, where ${\vec u^i}\equiv \hbox{lim}_{\gamma\rightarrow 0} \delta {\vec R_{n.a.}^i}/\gamma$.  Since the linear approximation $\delta {\vec R_{n.a.}^i}= \gamma {\vec u^i}$ is not exact, there are small residual forces on the nodes, but an iterative perturbative procedure can be used to determine the correction to the leading order result.  These residual forces may be estimated following Pythagoras' theorem: the transverse relative displacement at a contact cause a strain and therefore a residual force of the order of $\delta R^\bot_{n.a.}{}^2(\gamma)\sim \gamma^2 \delta z ^{2\phi}$. When this quantity becomes larger that  ${\vec F^i}\sim \gamma$, the linear approximation breaks down. This occurs for some $\gamma^*\sim \delta z^\beta\sim \delta z^{-2\phi}$, yielding the relation $\beta=-2\phi=1$ observed in our numerical simulations. The resulting divergence in the particle velocity explains why non-linearities occur at a very small strain close to the rigidity onset. 

The previous argument also yields the scaling form for the dimensionless shear stress $G$ in the floppy regime. Indeed when a shear strain $\gamma$ is imposed, each weak spring stores a dimensionless energy of order  $\delta E \sim k_w \delta {V^\bot_{n.a.}}^2$, leading to  $G\sim \rho_w k_w \delta E/ \gamma^2 \sim \rho_w k_w/|\delta z|$ in the floppy regime, as observed in Fig(\ref{fig3}.b). Obviously, this scaling  is expected to fail near the rigidity threshold, and we expect a cross-over to occur when $G\sim \delta z$ as expected for the backbone from our earlier scaling arguments. This allows us to define a characteristic coordination scale $u^*\sim \sqrt{\rho_wk_w}$ so that there is an associated critical strain $\gamma^*  \sim u^*\sim \sqrt{\rho_wk_w}$ where our two estimates are of the same order. This defines three regimes for the mechanical response as a function of the relative coordination: (i) For $\delta z << -u^* $,  $G\sim (\rho_w k_w)/|\delta
z|$ and $\gamma^* \sim - \delta z$. The rigid backbone is not significantly deformed. (ii) For  $-u^*<< \delta z< < u^*$, the energy is
shared between strong and weak interactions, $G \sim  u^* \sim \sqrt{\rho_w k_w }$ and $\gamma^* \sim u^*$. (iii) For $\delta z >> u^*$,  most of the energy is
condensed in the rigid backbone, $G\sim \delta z$ and $\gamma^* \sim \delta z$. Fig(\ref{fig3}.c) shows the
behavior of the shear modulus as the coordination increases, and the cross-over from the floppy to the rigid
network. To confirm   this description, we compute $\delta V^\bot_{n.a.}$ for different coordinations. The result is plotted in Fig(\ref{fig3}.d) for $k_w=1/300$. Although the ratio of stiffnesses between weak and strong springs is large, the intermediate region is of significant amplitude and
vanishes only as the square root of this ratio, due to the "lever" effect induced by the backbone.

We conclude with a brief discussion of our results and its implications. Using numerical simulations of a weakly coordinated network as an exploratory tool, we have shown that  (i) rigid and floppy networks are controlled by the same critical point (ii) two exponents, $\phi$ and $\theta$ which characterize the amplitude of non-affine displacements and the shear modulus respectively completely characterize the system; all other exponents describing the effects of non-linearities and  the stiffness induced by weak interactions follow from these. Furthermore, near the rigidity threshold, the amplitude of the non-affine displacement  rapidly increases and the material response is then characterized by a point on a two-dimensional phase diagram  $(\delta z, u^*)$. In the context of glasses, our model may describe tetrahedral network glasses, such as silica, amorphous silicon or water, where fluctuations of coordination are rare at reasonable pressures. Such networks,  where the joints linking tetrahedra are soft, are marginally rigid \cite{dove,thesis}, and the dominant weak interaction
rigidifying the system is the energy required to change the angle between two adjacent tetrahedra, whose strength is small for silica, and much larger for silicon. Thus silica is expected to behave effectively as a weakly-coordinated network, and exhibit large non-affine displacements, while amorphous silicon should not.  In contrast, for chalcogenide glasses where the composition of atoms of different valence can be modified to control rigidity \cite{phillips}, the precise topology of the covalent network near the rigidity threshold is  still unsettled \cite{bool,tan2}. Our model is presumably the simplest approximation of such networks and  gives a plausible explanation for the observation of a smooth cross-over in the elastic moduli near $z_c$ \cite{baza}: although Van der Waals interactions are two orders of magnitude softer than covalent bonds, their effect is not negligible near $z_c$ due to the lever effect of the backbone that we have described, and further predicts that the non-affine displacement will be maximum near the rigidity threshold. In  solid foams or stiff fiber networks, the different interactions at play (bending, stretching, cross-link rigidity...)  cause a change in the effective coordination. Since bending is a softer mode than stretching, those fibrous systems will be in general floppy.  Our work  yields the simple prediction that the strain $\gamma^*$ at which the system begins to stiffen and the amplitude of the non affine displacement will be anti-correlated, and that $\gamma^*$ will decrease as cross-links are added or fiber length is increased.
At a quantitative level, we expect the corresponding exponents  to depend on the particular structural properties of the network \cite{levine,frey}. In particular,  (i) if long fibers are present, non-affine displacements are enhanced, $\delta V^\bot_{n.a.}\sim1/|\delta z|$ \cite{frey}, and non-linearities are expected to occur at a smaller strain, and  (ii) if the link size is widely distributed, small links will effectively act as points of higher coordination, which  will also tend to reduce $\gamma^*$. More generally, our study suggests that the amplitude of the non-affine displacements allows us to classify amorphous solids according to their closeness to criticality in the coordination number.

\begin{acknowledgments} 
\vspace*{-0.3cm}
We thank Ning Xu for providing the jammed configurations and Oskar Hallatscheck for comments on the manuscript. 
\end{acknowledgments}

\end{document}